\documentclass[epj,final]{svjour}

\usepackage{latexsym}
\usepackage{url}
\usepackage{amsfonts}

\RequirePackage{graphicx}

\begin{document}
\title{Towards investigation of evolution of dynamical systems with independence of time accuracy: more classes of systems}
\author{V. G. Gurzadyan\inst{1,2},  A.A. Kocharyan\inst{2,3}
}                     
%
%
\institute{SIA, Sapienza University of Rome, Rome, Italy \and Center for Cosmology and Astrophysics, Alikhanian National Laboratory and Yerevan State University, Yerevan, Armenia \and 
School of Mathematical Sciences, Monash University, Clayton, Australia}
\date{Received: date / Revised version: date}
%

\abstract{
The recently developed method (Paper 1) enabling one to investigate the evolution of dynamical systems with an accuracy not dependent on time is developed further. The classes of dynamical systems which can be studied by that method are much extended, now including systems that are; (1) non-Hamiltonian, conservative; (2) Hamiltonian with time-dependent perturbation; (3) non-conservative (with dissipation). These systems cover various types of N-body gravitating systems of astrophysical and cosmological interest, such as the orbital evolution of
planets, minor planets, artificial satellites due to tidal, non-tidal perturbations and thermal thrust, evolving close binary stellar systems,
and the dynamics of accretion disks.
} 

\maketitle
\section{Introduction}

The typical procedure used in simulations of N-body gravitating systems \cite{A} is the numerical iterative integration of the equations of motion, but this has a well-known difficulty: the inevitable accumulation of errors as the number of iterations increases and hence beginning from a distinct time further computations become meaningless. The difficulty is serious in the case of typical nonlinear and/or many dimensional systems covering, as now is understood, most of the physical systems. The systems of N gravitating bodies involving various astrophysical problems possess chaotic properties (e.g. \cite{GS,GK1,GK2,GK3,LR,L}), which make their long term simulations rather complex.

Recently a method of numerical investigation was developed \cite{GHK}, providing a principal possibility for determination of the evolution of dynamical systems with an accuracy not dependent on time: an example of an evolution of each parameter of a system at
given accuracy was exhibited in \cite{GHK}. This method was here shown to be applicable for Hamiltonian systems with small periodic perturbation not dependent on time \cite{Arn}

\begin{equation}
H(I,\vartheta,\beta) =H_0(I)+\beta H_1(I,\vartheta)\,, 
\end{equation}
where $H_0(I)$ is the integrable Hamiltonian, $I, \vartheta$ are the action-angle variables and $\beta$ is the small parameter
of the non-integrable Hamiltonian $H_1$.

Here we represent a further development of that method extending the class of dynamical systems which can be investigated, paying mainly attention to the systems having primary physical and astrophysical interest. We show that systems of rather general classes can be principally studied, including systems that are:

1. non-Hamiltonian, conservative;

2.	non-conservative (with dissipation);

3.	Hamiltonian, with time-dependent perturbation.

In our analysis we essentially use a result by Chernoff \cite{Ch} concerning the properties of operators on a complete Riemannian manifold.

Our initial aim is to investigate a rather general type first order differential equations
\begin{equation}
\dot{x}^a = f^a(x)\, .	
\end{equation}

The main idea of the method developed in \cite{GHK} was the reduction of this equation to the following one
\begin{equation}
i\dot u + Lu = 0\, ,
\end{equation}
where $L$ is a differential operator. The solution of this equation by means of the computation of the resolvent using computer algebraic analytical codes gives the evolution of the function $u(t)$ in time. Then the functions $x^a(t)$, being determined by the function $u(t)$ and describing the initial system, can be found (see \cite{GHK}).

\section{The resolvent}

Let	$M$	be a $d$-dimensional complete Riemannian manifold with a metric $g$.	Denote by
$F(M)$ the set of	smooth functions determined on $M$. Let $F_0g(M)$ be the subset of $F(M)$ which
includes functions having a compact support. On $F_0(M)$ one has the operation of inner product
\begin{equation}
(u,v) = \int_M u^*(x)v(x)(\det g)^{1/2}d^d x\, ;    
\end{equation}
where $u$ and $v$ are elements of $F_0(M)$. Denote by $L^2(M)$ the completion of $F_0(M)$ with respect to that product.

Consider the operator
$$
L : F(M) \Rightarrow F(M)
$$
in the form

\begin{equation}
Lu=-i \sum^d_{a=1}f^a(x)\frac{\partial u}{\partial x} a \equiv -if^a\partial_a u;
\end{equation}
where $f^a(x)$ are smooth functions. 

One can show that the following relation for this operator is fulfilled \cite{Arn}

$$
(u, Lv) = \int_M u^*(x)Lv(x)(\det g)^{1/2}d^d x
$$
\begin{equation}   
=i\int_Mu^*(x)v(x)(div f)(\det g)^{1/2}d^dx+(Lu,v)\, ,  
\end{equation}

where $u,v \in F_0(M)$,
$$
div f=(\det g)^{-1/2}\partial_a((\det g)^{1/2}f^a)=f^a; a\, .    
$$

When
\begin{equation}
div f = 0\, ,
\end{equation}
then
\begin{equation}
(u,Lv) = (Lu,v)\, ,
\end{equation}
i.e. $L$ is a formal self-adjoint operator.

In our case the velocity of the propagation $c(x)$ introduced by Chernoff \cite{Ch} is equal to

\begin{equation}
c(x) = \left\| f(x)\right\| = (g_{ab}f^a f^b)^{1/2}\, . 
\end{equation}

According to Chernoff (Theorem 2.2 in \cite{Ch}), if $M$ is a complete Riemannian manifold and
$$
\int^\infty dr/c(r)=+\infty\, ,
$$ 
where $c(r) = \sup\{c(x); x \in S(y,r)\}, S(y,r)$ is a ball of radius $r$ with center $y$, then the operator $L$ defined on space $L^2(M)$ with domain $F_0(M)$ is essentially self-adjoint.

Consider now Eq.(3), to which, as mentioned above, Eq.(2) is reduced.


The solution of that equation can be found by means of the Laplace transform \cite{GHK},  
\begin{equation}
u(x,t)=-\int_\Gamma\exp(-i\lambda t)R_\lambda(\beta)u(x,0)\frac{d\lambda}{2\pi i}\, ;
\end{equation}
\begin{equation}
\Gamma=\{\sigma+is\, , -\infty<\sigma<+\infty\, , s=s_0>0\}
\end{equation}
where $R_\lambda (\beta)$ is the resolvent,
\begin{equation}
R_\lambda(\beta)=(\lambda-L)^{-1}\, , Im(\lambda)>0\, .
\end{equation}

Let us first consider the case when

\begin{equation}
f^a(x)=\xi^a+\beta\zeta^a(x)
\end{equation}
and

\begin{equation}
div \xi=0= div \zeta,\, \beta = const \ll 1\, , 
\end{equation}
then the resolvent is well approximated by formula \cite{GHK}

\begin{equation}
R_\lambda(\beta) = \sum ^N_{k=0}\beta^k(R_\lambda(0)B)^k R_\lambda(0)+o(\beta^N)\, ,
\end{equation}
where
$$
R_\lambda=(\lambda-L_0)^{-1}\, ,
$$
$$
L_0=-i\xi^a\partial_a\, ,
$$
$$
B=-i\zeta^a\partial_a\, .
$$

In the case when for the dynamical system defined on $M$

\begin{equation}
div f \neq 0\, ,
\end{equation}
we shall consider another one, having in local coordinates the form
$$
\dot{x}^a = f^a(x)\, ,
$$
$$
\dot{y} = -y\, div\, f\,.
$$
Here the manifold $M$ is reduced to $M\times R$ with the metric
\begin{equation}
g + dy^2\, .
\end{equation}

In the particular case that

\begin{equation}
M^d = Tor^k \times R^{d-k} = \{\vartheta^1,\ldots ,\vartheta^k\}\times\{I_1,\ldots,I_{d-k}\} 
\end{equation}
and
$$
\dot{\vartheta}^a=\xi^a(I), a=1,\ldots,k\,,
$$
$$
I^b=0, b=1,\ldots,d-k\, ,
$$
the resolvent $R_\lambda(\beta)$ can be found as in \cite{GHK}.

\section{Conclusions}

Let us summarize the dynamical systems covered by the above analysis. 

The systems represented by equations
$$
\dot{x}^a=f^a(x)\, ,
$$
include:

1.	Systems for which the Liouville theorem is fulfilled,
$$
div f = 0\, .
$$

This problem is reduced to finding out of the resolvent $R_\lambda(0)$, including the case when:

a)	the numbers of action-angle variables do not equal to each other. The calculations of $R_\lambda(0)$ should be considered separately at each given case. For example, $R_\lambda(0)$ is easily found and therefore the method is trivially applicable when

b)	$f^a$ is a periodic function;

2.	Systems with perturbed Hamiltonian

\begin{equation}
H(I,\vartheta,t,\beta)=H_0(I)+\beta H_1(I,\vartheta,t).\, 
\end{equation}

When $H(I,\vartheta, t,\beta)$ is a periodic function of time the resolvent is trivially represented via Fourier series (cf. \cite{GHK}).
In other cases the $R_\lambda(0)$ must be calculated separately;

3.	Non-conservative systems, when
$$
div f \neq 0\, .
$$

This case is reduced to P.1, as shown above.

The evolution of all these dynamical systems can be investigated by the method developed in \cite{GHK} with an accuracy not dependent on the value of time. Evidently, now the classes of physical and astrophysical problems are far extended. Namely,  besides traditional astrophysical nonlinear dynamical systems - N-body gravitating systems - now this will include an entire bunch of problems, e.g. the non-Hamiltonian motion of minor planets and space probes at tidal, non-tidal perturbation modes, and thermal thrust; these are particularly important, for example, for accurately testing of frame dragging, e.g.\cite{GC}. Other examples of non-conservative dissipative systems include e.g. the orbital evolution of close binary stellar systems due to evolving tidal effects, the magnetohydrodynamical evolution of non-stationary accretion disks near black holes, see e.g. \cite{LBP,GO,OB}.

\end{document}